\documentclass{article}
\usepackage{graphicx,amssymb,amsmath,times}

\def \gcm {$\mathrm{\ g\ cm^{-2}}$}

\def \eev { $\mathrm{\ EeV}$}

\def \deg {$^\circ$}

\def \pao {Pierre Auger Observatory }

\begin{document}

\title{Upper limit on the diffuse flux of UHE tau neutrinos \\
from the Pierre Auger Observatory}

\maketitle

\begin{center}
{\bf The Pierre Auger Collaboration} \\
\end{center}

\par\noindent
J.~Abraham$^{14}$, 
P.~Abreu$^{69}$, 
M.~Aglietta$^{55}$, 
C.~Aguirre$^{17}$, 
D.~Allard$^{33}$, 
I.~Allekotte$^{7}$, 
J.~Allen$^{89}$, 
P.~Allison$^{91}$, 
J.~Alvarez-Mu\~{n}iz$^{76}$, 
M.~Ambrosio$^{58}$
% ...

\par\noindent
\\

% ...
$^{7}$ Centro At\'{o}mico Bariloche, Comision Nacional de Energ\'{\i}a 
At\'{o}mica and Instituto Balseiro (CNEA-UNC), San Carlos de 
Bariloche, Argentina \\
% ...
$^{14}$ Universidad Tecnol\'{o}gica Nacional, FR-Mendoza, Argentina
 \\
% ...
$^{17}$ Universidad Catolica de Bolivia, La Paz, Bolivia \\
% ...
$^{33}$ Laboratoire AstroParticule et Cosmologie, Universit\'{e} 
Paris 7, IN2P3/CNRS, Paris, France \\
% ...
$^{55}$ Istituto di Fisica dello Spazio Interplanetario (INAF),
 Universit\`{a} di Torino and Sezione INFN, Torino, Italy \\
% ...
$^{58}$ Sezione INFN di Napoli, Napoli, Italy \\
% ...
$^{69}$ LIP and Instituto Superior T\'{e}cnico, Lisboa, Portugal \\
% ...
$^{76}$ Universidad de Santiago de Compostela, Spain \\
% ...
$^{89}$ New York University, New York, NY, USA \\
% ...
$^{91}$ Ohio State University, Columbus, OH, USA \\
% ...


\newpage

\begin{abstract}
The surface detector array of the Pierre Auger Observatory is
sensitive to Earth-skimming tau-neutrinos $\nu_\tau$ that interact in
the Earth's crust.  Tau leptons from $\nu_\tau$ charged-current
interactions can emerge and decay in the atmosphere to produce a
nearly horizontal shower with a significant electromagnetic component.
The data collected between 1 January 2004 and 31 August 2007 are used
to place an upper limit on the diffuse flux of $\nu_\tau$ at EeV
energies. Assuming an $E_\nu^{-2}$ differential energy spectrum the
limit set at 90 $\%$ C.L. is
$E_\nu^{2}~\mathrm{d}N_{\nu_\tau}/\mathrm{d}E_{\nu} < 1.3 \times
10^{-7}$ GeV cm$^{-2}$ s$^{-1}$ sr$^{-1}$ in the energy range
$2\times10^{17} \mathrm{eV} < E_\nu < 2\times10^{19}$~eV.
\end{abstract}


\maketitle

\vline{}

The detection of Ultra High Energy (UHE) cosmic neutrinos at EeV (1
EeV $\equiv 10^{18}~$eV) energies and above is a long standing
experimental challenge. Many experiments are
searching for such neutrinos, and there are several ongoing
efforts to construct dedicated experiments to detect
them~\cite{Halzen:2002pg,Halzen:2007ip,Falcke:2004aw}. Their discovery
would open a new window to the universe~\cite{Becker:2007sv}, and
provide an unique opportunity to test fundamental particle physics at
energies well beyond current or planned accelerators. The observation
of UHE Cosmic Rays (UHECRs) requires that there exist UHE cosmic
neutrinos, even though the nature of the UHECR particles and their
production mechanisms are still uncertain. All models of UHECR origin
predict neutrino fluxes from the decay of charged pions which are
produced either in interactions of the cosmic rays in their sources,
or in their subsequent interactions with background radiation
fields. For example, UHECR protons interacting with the Cosmic
Microwave Background (CMB) give rise to the so-called `cosmogenic' or
GZK neutrinos~\cite{GZKnus}. The recently reported suppression of the
cosmic ray flux above $\sim4 \times
10^{19}~$eV~\cite{Abbasi:2007sv,Yamamoto:2007xj,Collaboration:2007ba} as 
well as the observed correlation of the highest energy cosmic rays with 
relatively nearby extragalactic objects \cite{Collaboration:2007bb} both 
point to UHECR interactions on the infrared or microwave backgrounds during
extragalactic propagation. These interactions must result in UHE
neutrinos although their flux is somewhat uncertain since this depends
on the primary UHECR composition and on the nature and cosmological
evolution of the sources as well as on their spatial
distribution~\cite{Engel:2001hd,Allard:2006mv}.
 
Tau neutrinos are suppressed in such production processes relative to
$\nu_e$ or $\nu_\mu$, because they are not an end product of the
charged pion decay chain and far fewer are made through the production
and decay of heavy flavours such as charm. Nevertheless, because of
neutrino flavour mixing, the usual 1:2 ratio of $\nu_e$ to $\nu_\mu$ at
production is altered to approximately equal fluxes for all flavours
after travelling cosmological distances~\cite{Learned:1994wg}. Soon
after the discovery of neutrino oscillations~\cite{Fukuda:1998} it was
shown that $\nu_{\tau}$ entering the Earth just below the horizon
(Earth-skimming)~\cite{Fargion:2000iz,LetessierSelvon:2000kk,Feng:2001ue}
can undergo charged-current interactions and produce $\tau$ leptons.
Since a $\tau$ lepton can travel tens of kilometers in the Earth at EeV
energies, it can emerge into the atmosphere and decay in flight
producing an nearly horizontal extensive air shower (EAS) above the
detector. In this way the effective target volume for neutrinos can be
rather large.

The \pao \cite{Abraham:2004dt} has been designed to measure UHECRs with
unprecedented precision. Detection of UHECRs is being achieved exploiting
the two available techniques to detect EAS, namely, arrays of surface
particle detectors and telescopes that detect fluorescence
radiation. UHE particles such as protons or heavier nuclei interact
high in the atmosphere, producing showers that contain muons and an
electromagnetic component of electrons, positrons and
photons. This latter component reaches a maximum at an atmospheric
depth of order 800\gcm, after which it is gradually attenuated. 
Inclined showers that reach the ground after travelling
through 2000\gcm$\mathrm{\ }$ or more of the atmosphere are dominated
by muons arriving at the detector in a thin and flat shower front.

The surface detector (SD) array of the Pierre Auger Observatory can
be used to identify neutrino-induced
showers~\cite{Capelle:1998zz,Bertou:2001vm,Zas:2005zz}. The
fluorescence detectors can also be used for neutrino
searches~\cite{Aramo:2004pr,Miele:2005bt} but the nominal
$10\%$ duty cycle of the fluorescence technique reduces the
sensitivity. The electromagnetic component of neutrino-induced showers
might reach the ground if the shower develops close enough to the
detector, producing a signal which has a longer time duration than for an 
inclined shower initiated by a nucleonic primary. Thus
close examination of inclined showers enables showers developing near
to the ground and those produced early in the atmosphere to be
distinguished. This allows the clean identification of showers induced
by neutrinos, and in particular those induced by $\nu_{\tau}$, with
the SD~\cite{Billoir:2007zz,blanchICRC07,JaimeICRC07}.

Here we present the result of a search for deep, inclined, showers in
the data collected with the SD of the Pierre Auger Observatory. 
Identification criteria have been developed to find EAS that are 
generated by $\tau$ leptons emerging from the Earth. No candidates have 
been found in the data collected between 1 January 2004 and 31 August 
2007 --- equivalent to roughly one year of operation of the planned
full array.

The construction of the Southern \pao in Mendoza, Argentina, is currently close
to being completed. It consists of an array of water Cherenkov tanks
arranged in a hexagonal grid of 1.5~km covering an area of 3000~km$^2$ that 
is overlooked by 24 fluorescence telescopes located at four sites around
the perimeter. The array comprises 1600 cylindrical tanks of 10~m$^2$ surface 
containing purified water, 1.2~m deep, each  
instrumented with $3 \times 9''$ photomultiplier tubes sampled by 40~MHz Flash
Analog Digital Converters (FADCs)\cite{Abraham:2004dt}. Each tank is regularly 
monitored and calibrated in units of Vertical Equivalent Muon (VEM)
corresponding to the signal produced by a $\mu$ traversing the tank 
vertically~\cite{Bertou:2005ze}.

The procedure devised to identify neutrino candidate events within the
data set is based on an end-to-end simulation of the whole process,
from the interaction of the $\nu_{\tau}$ inside the Earth to the
detection of the signals in the tanks. The first step is the
calculation of the $\tau$ flux emerging from the Earth. This is done
using a simulation of the coupled interplay between the $\tau$ and the
$\nu_\tau$ fluxes through charged-current weak-interactions and $\tau$
decay, taking into account also the energy losses due to neutral
current interactions for both particles, and bremsstrahlung, pair
production and nuclear interactions for the $\tau$ lepton. The
emerging $\tau$ flux can be folded with the $\tau$ decay probability to give
the differential probability of $\tau$ decaying in the atmosphere as a
function of its energy and decay altitude,
$\mathrm{d}^{2}p_\tau/\mathrm{d}E_\tau\mathrm{d}h_c$.

Modelling of the showers from $\tau$ decays in the atmosphere is
performed using the AIRES code~\cite{Aires260}. The TAUOLA
package~\cite{Jadach:1993hs} is used to simulate $\tau$ decay and
obtain the secondary particles and their energies. Showers induced by
the products of decaying $\tau$s with energies between $10^{17}$ to $3
\times 10^{20}$ eV are simulated at zenith angles ranging
between $90.1^\circ$ and $95.9^\circ$ and at an altitude of the decay
point above the Pierre Auger Observatory in the range $0-2500$
m. Finally, to evaluate the response of the SD to such events,
the particles reaching the ground in the simulation are stored and
injected into a detailed simulation of the SD~\cite{Ghia:2007fh}.

\begin{figure}
\includegraphics[width=0.90\textwidth]{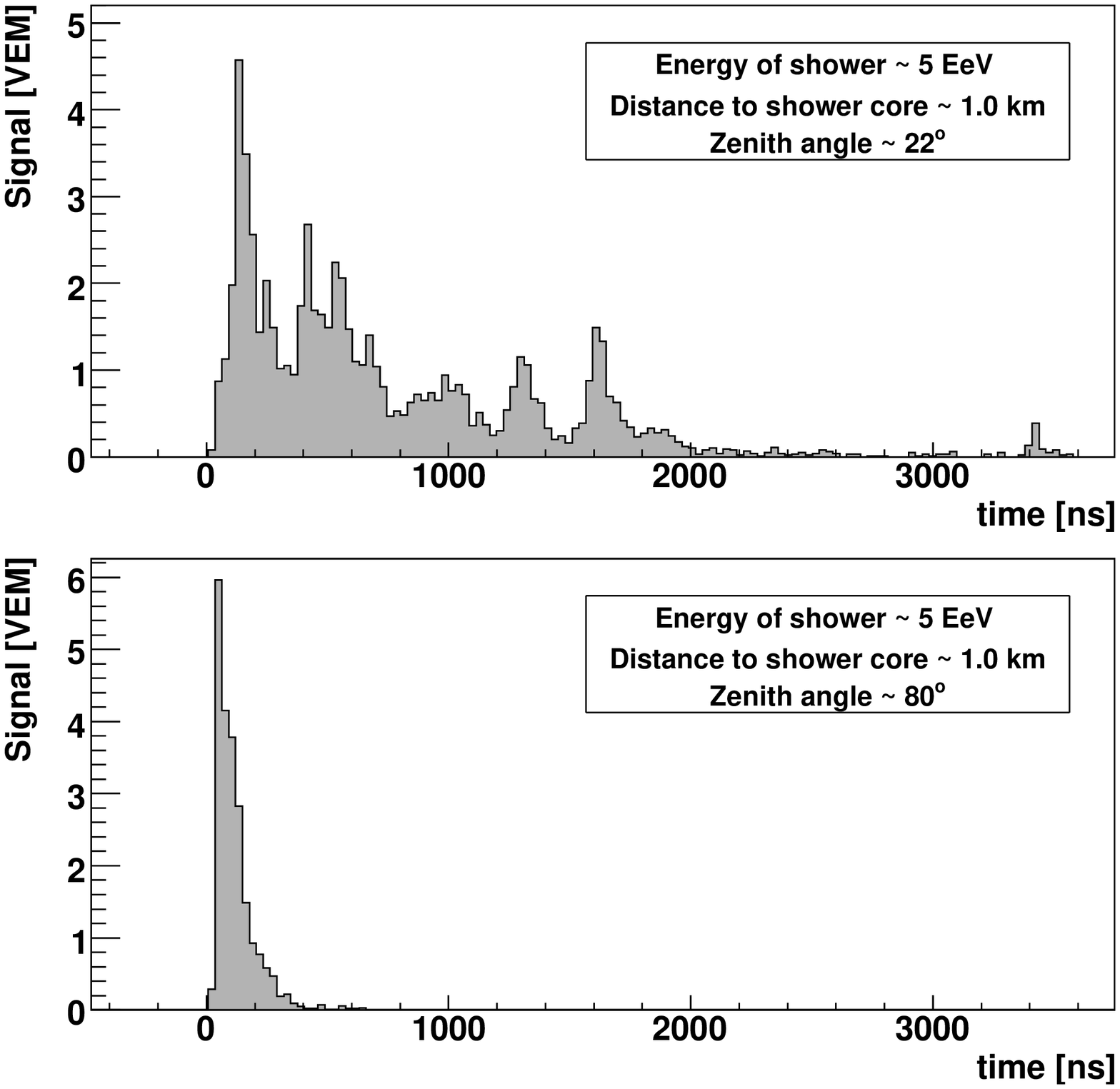}
\caption{\label{fig1}FADC traces of stations at 1 km from the shower
core for two real showers of 5\eev. Top panel: electromagnetic
component ($\theta\sim$ 22 \deg); bottom: muonic signal ($\theta\sim$
80 \deg).}
\vspace{-0.35cm}
\end{figure}

\begin{figure*}
\begin{center}
\includegraphics [width=0.90\textwidth]{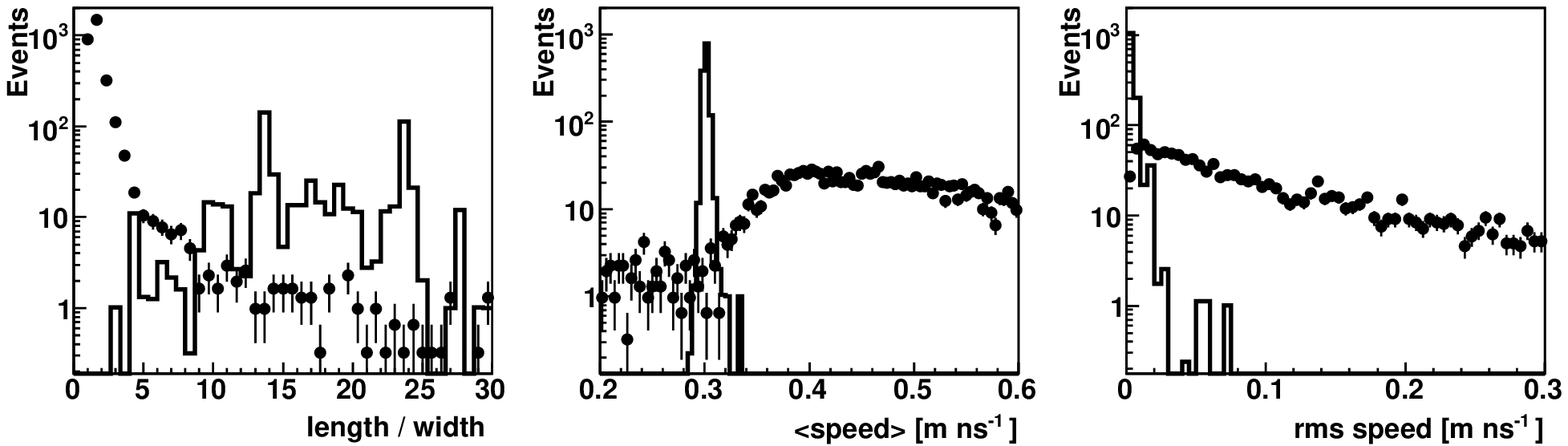}
\vspace{-0.35cm}
\end{center}
\caption{Distribution of discriminating variables for showers
initiated by $\tau$s decaying in the atmosphere, generated by
$\nu_\tau$s with energies sampled from an $E_\nu^{-2}$ flux
(histogram), and for real events passing the ``young shower''
selection (points). Left: length/width ratio of the footprint of the
shower on the ground; middle: average speed between pairs of stations;
right: r.m.s. scatter of the speeds. See text for details.}
\label{fig2}
\vspace{-0.35cm}
\end{figure*}

A set of conditions has been designed and optimized to select showers
induced by Earth-skimming $\nu_{\tau}$, rejecting those induced by
UHECR. The 25 ns time resolution of the FADC traces allows unambiguous
distinction between the narrow signals induced by muons and the broad
signals induced by the electromagnetic component (Figure~\ref{fig1}). For
this purpose we tag the tanks for which the main segment of the FADC
trace has 13 or more neighbouring bins over a threshold of 0.2 VEM,
and for which the ratio of the integrated signal over the peak height
exceeds 1.4. A neutrino candidate is required to have over $60\%$ of
the triggered tanks satisfying these ``young shower" conditions as well as 
fulfilling the central trigger condition~\cite{Abraham:2004dt} with these
tanks. In addition the triggered tanks are required to have elongated
patterns on the ground defining the azimuthal arrival direction (as
expected for inclined events) by assigning a length and a width to the
pattern and restricting its ratio (length/width$>$5). Finally, we calculate the apparent
speed of the signal moving across the ground along the azimuthal
direction, using the arrival times of the signals at ground and the
projected distances between tanks. The average speed, as measured
between pairs of triggered stations, is required to be compatible with
that expected for an event traveling close to the horizontal direction by
requiring it to be very close to the speed of light, in the range
(0.29, 0.31) m ns$^{-1}$ with an r.m.s. scatter below $0.08~{\rm
m~ns^{-1}}$. These conditions are found to retain about 80~$\%$ of the
simulated $\tau$ showers triggering the SD. The final sample is
expected to be free of background from UHECR-induced
showers. In Figure \ref{fig2}, we show the distributions of these
discriminating variables for real events and simulated $\tau$ showers.

Over the period analyzed, no candidate events were found that fulfilled
the selection criteria. Based on this, the Pierre Auger Observatory data
can be used to place a limit on the diffuse flux of UHE $\nu_{\tau}$. 
For this purpose the exposure of the detector must be evaluated. The total 
exposure is the time integral of the 
instantaneous aperture which has changed as the detector has grown while 
it was being constructed and set into operation. 

Calculation of the effective aperture for a fixed neutrino energy $E_{\nu}$ 
involves folding the aperture with the conversion probability and the
identification efficiency.
The identification efficiency $\epsilon_{\rm ff}$ depends on the $\tau$
energy $E_\tau$, the altitude above ground of the central part of the
shower $h_c$ (defined at 10 km after the decay
point~\cite{Bertou:2001vm}), the position $(x, y)$ of the shower in
the surface $S$ covered by the array, and the time $t$ through the
instantaneous configuration of the array.
The expression for the exposure can be written as:
\begin{equation}
\mathrm{Exp} = \int_{\Omega} \mathrm{d}\Omega
\int_{0}^{E_\nu}\;\mathrm{d}E_\tau \int_{0}^{\infty}\;\mathrm{d}h_c
\frac{\mathrm{d}^2 p_\tau}{\mathrm{d}E_\tau \mathrm{d}h_c} B_\tau,
\label{Exposure2}
\end{equation}
\vspace{-0.3cm}
where
\begin{equation}
B_\tau (E_\tau, h_c) = 
\int_{T} \mathrm{d}t \int_{S} \mathrm{d}x \mathrm{d}y \, \cos\theta \, 
\epsilon_\mathrm{ff} [E_\tau, h_c, x, y, t]
\label{Acceptance}
\end{equation}
where $\theta$ and $\Omega$ are the zenith and solid angles.

The exposure is calculated using standard Monte Carlo techniques (MC) in two
steps. The first integral deals with the detector-dependent part,
including the time evolution of the array over the period $T$
considered (eq.\ref{Acceptance}). The integral in
$E_{\tau}$ and $h_c$ involves only the differential conversion
probability and $B_{\tau}$ (eq.\ref{Exposure2}). The estimated
statistical uncertainty for the exposure is below 3\%.

The MC simulations require some physical quantities that have not been
experimentally measured in the relevant energy range, namely the $\nu$
interaction cross-section, the $\tau$ energy loss, and the $\tau$
polarisation. The main uncertainty in these comes from the QCD structure
functions in the relevant kinematic range. We estimate the
uncertainty in the exposure due to the $\nu$ cross-section to be
15$\%$ based on the allowed range explored
in~\cite{Anchordoqui:2006ta}. The uncertainties in the $\tau$ energy
losses are dominated by the $\tau$ photonuclear cross section. The 40$\%$ 
difference among existing calculations for the $\tau$ energy 
losses~\cite{Bugaev:2003sw,Dutta:2005yt,Aramo:2004pr}, which use different 
structure functions, is used as the systematic uncertainty.
The two extreme cases of polarization give 30$\%$ difference in
exposure and we take this as the corresponding uncertainty. The
relevant range of the structure functions includes regions of
Bjorken-$x$ and squared 4-momentum transfer, $Q^{2}$, where no
experimental data exist. Only extrapolations that follow the
behaviour observed in the regions with experimental data have been considered. 

We also take into account uncertainties coming from neglecting the
topography around the site of the Pierre Auger
Observatory~\cite{Gora:2007nh} (18$\%$).
We adopt a 25$\%$ systematic uncertainty due to MC simulations of the
EAS and the detector, dominated by differences between hadronic models
(QGSJET~\cite{Kalmykov:1997te} and SIBYLL~\cite{Engel:1999db}).

\begin{figure*}
\begin{center}
\includegraphics [width=0.9\textwidth]{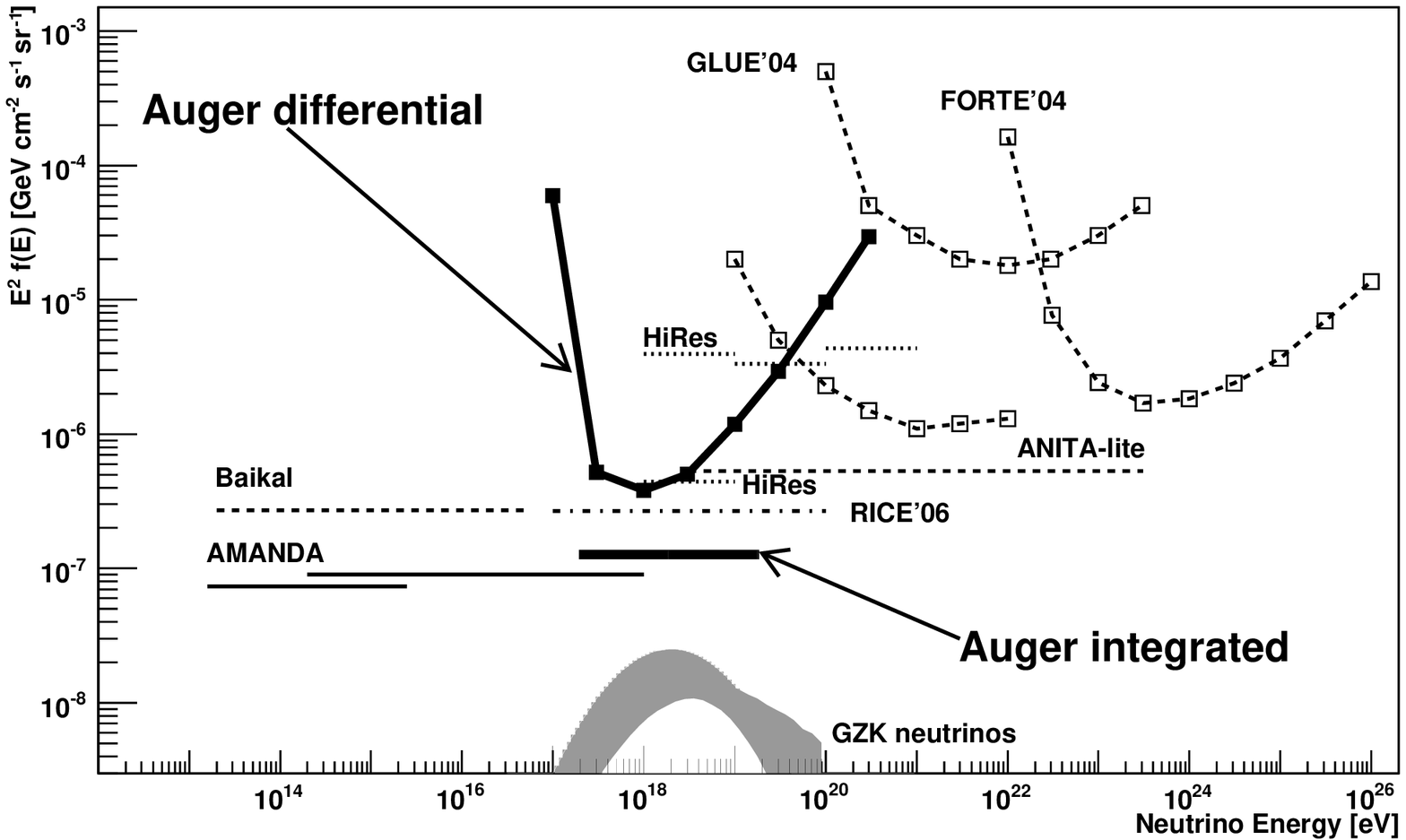}
\caption{Limits at 90$\%$ C.L. for a diffuse flux of $\nu_{\tau}$ from the 
  Pierre Auger Observatory. Limits from other
  experiments~\cite{Achterberg:2007qp,Ackermann:2007km,Martens:2007ff,Aynutdinov:2005dq,Kravchenko:2006qc,Barwick:2005hn,Gorham:2003da,Lehtinen:2003xv}
  are converted to a single flavour assuming a $1:1:1$ ratio of the 3
  neutrino flavours and scaled to 90$\%$ C.L. where needed. Two
  different formats are used: differential (squares) and
  integrated (constant lines). The shaded
curve shows the range of expected fluxes of GZK neutrinos from 
Ref.~\cite{Engel:2001hd,Allard:2006mv}, although 
predictions almost 1 order of magnitude lower and higher exist.}
\label{fig4}
\vspace{-0.35cm}
\end{center}
\vspace{-0.35cm}
\end{figure*}

Assuming a $f(E_\nu)\propto E_\nu^{-2}$ differential flux of $\nu_{\tau}$   
we have obtained a 90$\%$ C.L. limit on the diffuse flux of 
UHE $\nu_\tau$, whose level at 10$^{18}$~eV is representative for any
smooth spectral shape:
\begin{equation}
E_{\nu}^{2} f(E_\nu) < 1.0_{-0.5}^{+0.3} \times 10^{-7}
~{\rm GeV~cm^{-2}~s^{-1}~sr^{-1}} 
\label{limit}
\end{equation}
The central value is computed using the $\nu$ cross-section from
Ref.~\cite{Anchordoqui:2006ta}, the parametrisation of the energy
losses from Ref.~\cite{Dutta:2005yt} and an uniform random
distribution for the $\tau$ polarisation. The uncertainties correspond to
the combinations of systematic uncertainties in the exposure as given
above that lead to the highest/lowest neutrino event rate. The limit
is applicable in the energy range $2 \times 10^{17} - 2 \times
10^{19}$ eV, with a systematic uncertainty of about 15$\%$, over which 90$\%$ of the events are expected for
$f(E_\nu)\propto E_\nu^{-2}$. In Figure \ref{fig4}, we show our limit
adopting the most pessimistic scenario for systematic
uncertainties. It improves by a factor $\sim~3$ for the most
optimistic one. For energies above $10^{20}$~eV, limits are usually
quoted as $2.3/\mathrm{Exp} \times E_{\nu}$ for different energy
values (differential format), while at lower energies they are usually
given assuming an $E^{-2}$ flux (integrated format). We plot the
differential format to demonstrate explicitly that the sensitivity of
the Pierre Auger Observatory to Earth-skimming $\nu_\tau$ peaks in a
narrow energy range close to where the GZK neutrinos are expected.

The Earth-skimming technique used with data collected at the
surface detector array of the Southern Pierre Auger Observatory,
provide at present the most sensitive bound on neutrinos at EeV
energies. This is the most relevant energy to explore the predicted
fluxes of GZK neutrinos.
The Pierre Auger Observatory will continue to take data for about 20 
years over which time the limit should improve by over an order of magnitude
if no neutrino candidate is found.

\section*{Acknowledgments}
The successful installation and commissioning of the Pierre Auger Observatory
would not have been possible without the strong commitment and effort
from the technical and administrative staff in Malarg\"ue.

We are very grateful to the following agencies and organizations for financial support: 
Comisi\'on Nacional de Energ\'ia At\'omica, Fundaci\'on Antorchas,
Gobierno De La Provincia de Mendoza, Municipalidad de Malarg\"ue,
NDM Holdings and Valle Las Le\~nas, in gratitude for their continuing
cooperation over land access, Argentina; the Australian Research Council;
Conselho Nacional de Desenvolvimento Cient\'ifico e Tecnol\'ogico (CNPq),
Financiadora de Estudos e Projetos (FINEP),
Funda\c{c}\~ao de Amparo \`a Pesquisa do Estado de Rio de Janeiro (FAPERJ),
Funda\c{c}\~ao de Amparo \`a Pesquisa do Estado de S\~ao Paulo (FAPESP),
Minist\'erio de Ci\^{e}ncia e Tecnologia (MCT), Brazil;
Ministry of Education, Youth and Sports of the Czech Republic;
Centre de Calcul IN2P3/CNRS, Centre National de la Recherche Scientifique (CNRS),
Conseil R\'egional Ile-de-France,
D\'epartement  Physique Nucl\'eaire et Corpusculaire (PNC-IN2P3/CNRS),
D\'epartement Sciences de l'Univers (SDU-INSU/CNRS), France;
Bundesministerium f\"ur Bildung und Forschung (BMBF),
Deutsche Forschungsgemeinschaft (DFG),
Finanzministerium Baden-W\"urttemberg,
Helmholtz-Gemeinschaft Deutscher Forschungszentren (HGF),
Ministerium f\"ur Wissenschaft und Forschung, Nordrhein-Westfalen,
Ministerium f\"ur Wissenschaft, Forschung und Kunst, Baden-W\"urttemberg,
Germany; Istituto Nazionale di Fisica Nucleare (INFN),
Ministero dell'Istruzione, dell'Universit\`a e della Ricerca (MIUR), Italy;
Consejo Nacional de Ciencia y Tecnolog\'ia (CONACYT), Mexico;
Ministerie van Onderwijs, Cultuur en Wetenschap,
Nederlandse Organisatie voor Wetenschappelijk Onderzoek (NWO),
Stichting voor Fundamenteel Onderzoek der Materie (FOM), Netherlands;
Ministry of Science and Higher Education,
Grant Nos. 1 P03 D 014 30, N202 090 31/0623, and PAP/218/2006, Poland;
Funda\c{c}\~ao para a Ci\^{e}ncia e a Tecnologia, Portugal;
Ministry for Higher Education, Science, and Technology,
Slovenian Research Agency, Slovenia;
Comunidad de Madrid, Consejer\'ia de Educaci\'on de la Comunidad de Castilla
La Mancha, FEDER funds, Ministerio de Educaci\'on y Ciencia,
Xunta de Galicia, Spain;
Science and Technology Facilities Council, United Kingdom;
Department of Energy, Contract No. DE-AC02-07CH11359,
National Science Foundation, Grant No. 0450696,
The Grainger Foundation USA; ALFA-EC / HELEN,
European Union 6th Framework Program,
Grant No. MEIF-CT-2005-025057, and UNESCO.


\end{document}